# On the use of FHT, its modification for practical applications and the structure of Hough image

M. Aliev[1,3], E.I. Ershov[2], D.P. Nikolaev[2,3]

[1] Federal Research Center "Computer Science and Control" of Russian Academy of Sciences, Moscow, Russia

[2] The Institute for Information Transmission Problems of Russian Academy of Sciences, Moscow, Russia

[3]Smart Engines, Moscow, Russia

**ABSTRACT**

This work focuses on the Fast Hough Transform (FHT) algorithm proposed by M.L. Brady. We propose how to modify the standard FHT to calculate sums along lines within any given range of their inclination angles. We also describe a new way to visualise Hough-image based on regrouping of accumulator space around its center. Finally, we prove that using Brady parameterization transforms any line into a figure of type "angle".

**Keywords**: Hough transform, fast Hough transform, Hough-image, visualisation

## 1. INTRODUCTION

In the last decades interest in academic research for image analysis stays quite high. This is directly linked to the high pace of hardware development for image recording devices: photo cameras, scanners, X-ray machines, ultrasound machines, infrared cameras and others.

One of actively used means for image analysis is Hough Transform (HT) that allows to estimate the number of points lying near every line of certain set of lines. It is named after Paul Hough who described this transform in 1959 and obtained a patent in 1962 for "Method and means for recognizing complex patterns" [1] with the main method being the calculation of this transform on the image fragment. Since that time a lot of different algorithms were created that contain the reference to "Hough transform" in their name. They differ by the shape of the figure along the edges of which computation is performed and by the operation that is calculated along the figure (there are generalizations of HT for any associative-commutative operation).

At present the algorithm for calculating fast Hough Transform (FHT) is in high demand for solving a large number of practical problems either on PC or on low-powered embedded systems. FHT is used for finding page orientation angle [2], textual blocks rectification [3], detecting of edges (including document edges), barcode recognition, detection of roadway [4] and vanishing points[5], compensation of radial distortion [6]. Also FHT is used to accelerate adjustment of several binarization algorithms such as Otsu and Niblack [7]. Another use-case of FHT is computer tomography where FHT is a part of tomography reconstruction algorithm [8].

Algorithms, that use HT could be divided in two types:
1) Integral algorithms. Algorithms of this type use Hough transform as a discrete approximation for the continuous radon transform (more detailed link between HT and RT is described in [9]).
2) Combinatorial algorithms. Algorithms of this type have as their input a list of data (for example, points [10]) and calculate HT only on them (for example, only for lines going through any two input points).

Asymptotic of the first type algorithms depends on sizes of input image, while combinatorial algorithms asymptotic depends on the size of input array. This means that for algorithms of second type it is beneficial to perform image preprocessing for reduction of input data array, or for selecting it from the original image.

In this paper we review HT of the first type for discrete patterns of certain type [11] that approximate straight lines for two-dimensional image. This work is structured as follows: in Section 2 we overview existing parametrizations of HT. In Section 3 different aspects of HT are considered: padding of the original image for more convenient interpretation of Hough-image, obtaining full Hough-image, calculating coordinates of linear pattern from point in Hough-image. In Section 4 we describe the method of calculating sums along lines in a defined range of inclination angles on the original image and propose an algorithm modification for performing this computation. In Section 5 we present the operation "*fhtshift*" of regrouping Hough-image around center of accumulator and propose a method for its realization.

## 2. RELATED WORK

Hough suggested transform that maps points into lines and correspondingly collinear points to concurrent lines in order to analyse the resultant image[1]. Parametrization used by Hough was the equation of straight line with slope coefficient: point $(u, v)$ maps to line $x = vy + u$.

In 1972 Duda and Hart [12] pointed out that the method suggested by Hough is impractical because its result will not always belong to a bounded region. Later, in 2002 Bhattacharya, Rosenfeld and Weiss proved [13] that there is no such point-to-line mapping that transforms bounded area (original image) into bounded area (Hough image). To solve this problem they developed their own method which mapped points to sinusoid curves and correspondingly collinear point to concurrent curves and proved that the result of their transform with the parametrization suggested by them ($(\Theta, \rho)$ – angle between the perpendicular to the line and the abscissa axis 0X and its length) lies in a bounded region.

During the next years a lot of different types of parametrizations for calculating HT were created with the goal to improve the quality of line detection, reduce the calculation time or to make it more clear. As a result the consensus has been reached on which properties such a parametrization should have (uniqueness, boundness and uniformness) [14]. Also worth mentioning is the article referenced in the survey [15] that was written by R.Wallace in 1985 [16]. In this work the author suggests modified version of HT - so called muff-transform. In [17] authors propose to parametrize pattern with a point on the image and a certain kernel (the optimal formula for which is investigated in the work). Another interesting way of parametrization was suggested by Davies in [18] where a line is parametrized by the coordinates of the normal line base. In [19] authors demonstrate a method of the parallel parametrization [20] which usually serves for the visualization of multidimensional data. Coordinates in the original image are placed in correspondence with the set of equidistant parallel axis on each of them a point is set that corresponds to its coordinate. More detailed description of different parameterizations can be found in [15] and [9].

Despite the abundance of works on parametrizations, in most cases their efficiency was not sufficient and that led to development of different ideas for accelerating HT. Firstly in context of first approach it was suggested to calculate FHT using FFT [21]. In 1992 explicit scheme for FHT calculating was for the first time proposed by Brady and others [11] (though more known is his publication in 1998) "Approximate Discrete Radon Transform" (ADRT). Independently this algorithm was also invented in DEC in 1994 [22] as "Fast Linear Hough Transform" (FLHT) and in Innsbruck university in 1993 [23] ("Fast Discrete Radon Transform"). The advantage of the explicit scheme versus the method that uses FFT is in fact that there exist theoretical estimations for accuracy of line approximation with discrete patterns, and also it allows to use intermediate results of computations. However, at present the explicit scheme of FHT is proposed only for linear parametrization case. Therefore, it is necessary task to create tools for processing and analysis of Hough-image obtained with such parametrization.

## 3. ASPECTS OF HOUGH TRANSFORM

### 3.1. Padding the image by zero-filled fragment

For the following algorithm description to be clear we need some definitions that were first introduced in [11] and further developed in [2] and [7]. We consider an image $I$ with sizes $w \times h$ and line $p$. Let's define $\varphi$ as an angle between the perpendicular to $p$ from the origin O and the abscissa axis 0X, measured clockwise. All lines with $|tg(\varphi)| \leq 1$ we will now call "mostly vertical", and lines with $|ctg(\varphi)| \leq 1$ we will call "mostly horizontal". Let's consider the case where our line $p$ is "mostly vertical". In [2] the following parametrization is proposed: $(x_0, shift)$, where $x_0$ is the intersection of $p$ with the abscissa axis 0X, and $shift = h * tg(\varphi)$. The sign of $shift$ is fixed at positive so that negative shift is denoted as $-shift$. The case of "mostly horizontal" line $p$ is considered in the same way - the only difference in the parametrization is that we use the ordinate axis 0Y instead of the abscissa axis 0X (i.e. we take the intersection of $p$ with 0Y and shift along 0Y). For uniformity purposes, in case of "mostly horizontal" lines the original image is transposed.

Next, [7] defines the "quadrants" of angles for lines in the original image:
- $[-45°; 0°]$ (or $[315°; 0°]$) – "mostly vertical lines with positive shift"
- $[0°; 45°]$ – "mostly vertical lines with negative shift"
- $[45°; 90°]$ – "mostly horizontal with positive shift"
- $[90°; 135°]$ – "mostly horizontal with negative shift"

and shows that in Hough image for each quadrant every point of the original space corresponds to a line in the parameter space.

From now on without loss of generality we will consider only "mostly vertical" lines.

Some lines from the original image will intersect with the abscissa axis 0X outside the image (i.e. $x_0$ will be outside of the interval $[0; w-1]$). In this case computation of FHT will result in lines (which are transformed points from the original image) that "flow through" the boundaries of Hough-image, sometimes even creating a cyclic effect, which makes the analysis of Hough transform result more difficult (this violates duality of point and line). A way to resolve this difficulty by extending the original image by $h \times h$ is proposed in [11]. Since $shift = h * tg(\varphi) \leq h * tg(\frac{\pi}{4}) = h$ such an extension will increase the result of Hough transform to size $(w+h) \times h$. In case of "mostly horizontal" lines the original image is extended by $w \times w$, and the Hough transform result becomes $w \times (w+h)$.

### 3.2. Calculation of coordinates of linear pattern on the original image

Now let's describe the steps needed to obtain coordinates for linear pattern on the original image from a point $P(x_p, y_p)$ in Hough space. Again, we will consider the case of "mostly vertical" lines.

1) Calculate the coordinates of linear pattern $S$ which lies on the line $p(x_p, y_p)$ and connects the horizontal sides of the original image: for example, if we consider quadrant $[315°; 0°]$ then for point $(x_0, shift)$ this pattern will be from $(x_0, 0)$ to $(x_0 + shift, h)$.
2) If the calculated coordinates of both ends are on the edges of the original image then it's the pattern we were looking for, otherwise there are several options:
    a) if the calculated pattern intersects with the left side of the original image then we take part of the pattern that lies within the original image.
    b) if the calculated pattern crosses the right side of the extended image then (by the cyclic effect of the Hough transform) it will cross the left side of the original image. Therefore, we need to shift the pattern coordinates by $w+h$ along the abscissa axis 0X (in fact use the pattern coordinates by mod $(w+h)$) and this will bring us to the previous option (a)
    c) if the calculated pattern is contained fully in the "extension" part of the image it means that there is no pattern in the original image that corresponds to point $P$.

Illustration of these steps can be seen on figure 1.

It follows from (c) that Hough transform of such extended image will always (regardless of the input image) contain a triangular area with zero values (because there are no lines that have their images in this zone). We will call this area the "black zone" as opposed to the "meaningful" zone (the rest of the Hough image).

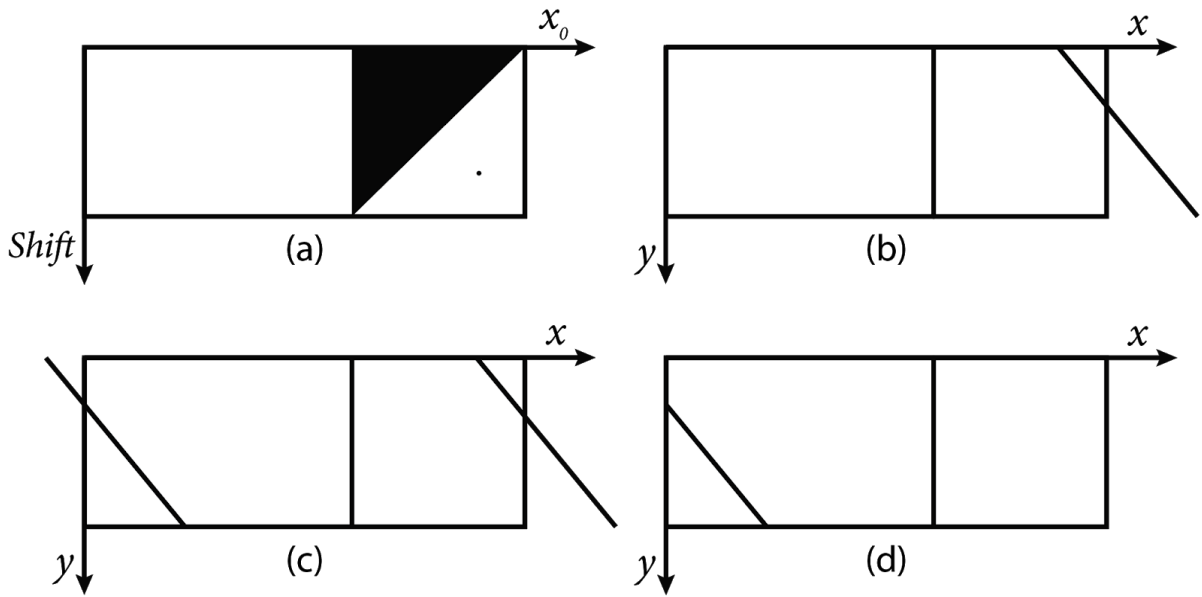

Figure 1 - (a) example of Hough transform for quadrant $[315°; 0°]$ in case of a completely white original image with one marked point; (b) result of the first calculation; (c) shifting the pattern along axis 0X; (d) final result

### 3.3 Full Hough transform

Performing Hough transform for only one range of angles (one quadrant) might not be enough for full analysis of the image. [7] suggests a method for concatenating the quadrants together one-by-one by their common edge rows (some quadrants will have to be flipped vertically): $[315°; 0°]$ flipped; $[0°; 45°]$; $[45°; 90°]$ flipped; $[90°; 135°]$.

Let's demonstrate that using this method preserves lines - i.e. images of points in Hough space remain straight linear pattern and not two connecting segments. Consider point $P(x_p, y_p)$. Its image for quadrant $[315°; 0°]$ will be the linear pattern between $(x_p, 0)$ and $(x_p - (h - y_p), h)$, equation for this line can be then written as: $shift = (\frac{h}{y-h}) * x_0 + (\frac{-h}{y-h}) * x_p$.

To concatenate the quadrants together we need to flip vertically the result of Hough transform, which changes the equation to $h - shift_{new} = (\frac{h}{y-h}) * x_0 + (\frac{-h}{y-h}) * x_p$, where $shift_{new}$ is the point coordinate on the concatenated and vertically flipped Hough-image. Rewriting it in standard form we have:

$$shift_{new} = (\frac{-h}{y-h}) * x_0 + (h + (\frac{h}{y-h}) * x_p) \quad (1)$$

Now let's consider the image of $P$ for quadrant $[0°; 45°]$: the linear pattern between $(x_p, 0)$ and $(x_p + (h - y_p), h)$. Here, the line equation will be: $shift = (\frac{h}{h-y}) * x_0 + (\frac{-h}{h-y}) * x_p$. In order to concatenate the quadrants together this one is shifted by $h$ along the axis $0X$, and the resulting equation is: $shift_{new} - h = (\frac{h}{h-y}) * x_0 + (\frac{-h}{h-y}) * x_p$. Again, rewriting it in standard form we have:

$$shift_{new} = (\frac{-h}{y-h}) * x_0 + (h + (\frac{h}{y-h}) * x_p) \quad (2)$$

We can see that (1) and (2) are identical and so the image of a point on the concatenated quadrants is indeed the same line.

The size of the concatenated together full Hough transform is $2 \times (w + h) - 3$. It is less than the total of individual results because each concatenation subtracts one row.

Example of original image and full Hough-image can be seen on figure 2.

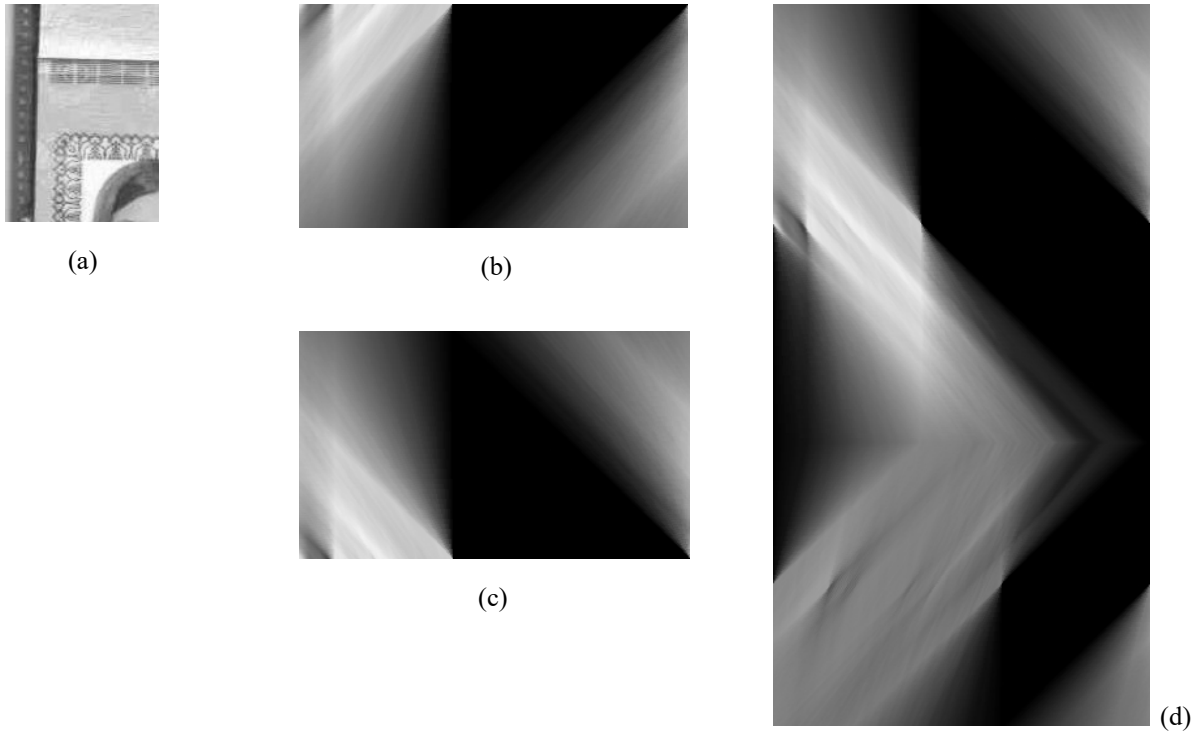

Figure 2 - (a) example of original image; (b) Hough transform for quadrant $[315°; 0°]$; (c) Hough transform for quadrant $[315°; 0°]$ after it was flipped vertically; (d) full transform result ($[315°; 135°]$).

When using full Hough transform to obtain the linear pattern we should first determine which quadrant contains our starting point. This can be done by its $shift$-coordinate since we know the order of concatenated quadrants: if $shift < h$ then the point lies in quadrant $[315°; 0°]$, if $h ≤ shift < 2 * h - 1$ (where $2 * h - 1 = h + (h - 1)$) is the total height of the first two quadrants in full Hough transform) then in $[0°; 45°]$, if $2*h-1 \leq shift < 2h+w-2$ then in quadrant $[45°; 90°]$ and finally if $2*h+w-2 \leq shift$ then $[90°; 135°]$. Now the point can be attributed to a particular quadrant; and we need to decrease the *shift*-coordinate by the total height of all previous quadrants.

## 4. WORKING WITH ARBITRARY ANGLE RANGE

Sometimes when working with an image we might need to perform a transform for an arbitrary range of angles $[\gamma 1; \gamma 2]$. This can be done in two steps:
1) Transforming the given range of angles into one analyzed previously (for which we already know how to perform Hough transform).
2) Perform Hough transform.

There are several operations that will allow us to complete the first step:
- Scale of original image along the abscissa axis 0X with scale factor $\alpha > 0$ (see figure 3). In this case angle $\beta = arctg\frac{HA}{h}$ changes into $\beta' = arctg(\frac{HA'}{h}) = arctg(\frac{\alpha HA}{h}) = arctg(\alpha * tg\beta)$. This means that quadrant $[-45°; 0°]$ changes into the range $[-arctg(\alpha); 0]$ and range $[-arctg(\frac{1}{\alpha}); 0]$ changes into $[-45°; 0°]$. Similarly, quadrant $[0°; 45°]$ will become $[0; arctg(\alpha)]$ and $[0; arctg(\frac{1}{\alpha})]$ will become $[0°; 45°]$.

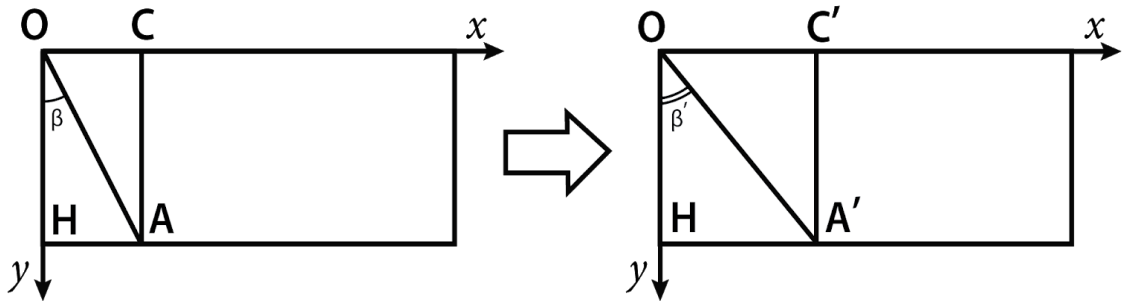

Figure 3 - example of scale with factor $\alpha = 2$, with point A changing into A'.

Thus, for any $0 < \beta < \frac{\pi}{2}$ the range $[-\beta; \beta]$ can be transformed into $[-45°; 45°]$. Applying the same operation to "mostly horizontal" case we can see that a range $[90°-\beta; 90°+\beta]$ can be transformed into $[45°; 135°]$.
- Horizontal shear of the original image by shear angle $\beta < \frac{\pi}{2}$ (see figure 4). In this case angle $\alpha = arctg\frac{HB}{h}$ changes into $\beta + \alpha' = arctg(\frac{A'H'+H'B'}{h}) = arctg(\frac{A'H'}{h} + \frac{H'B'}{h}) = arctg(tg\beta + tg\alpha)$. Similarly, the range $[-arctg(1-tg(\beta)); \beta]$ will become $[315°; 0°]$ (in this case line OB has angle $-\alpha$).

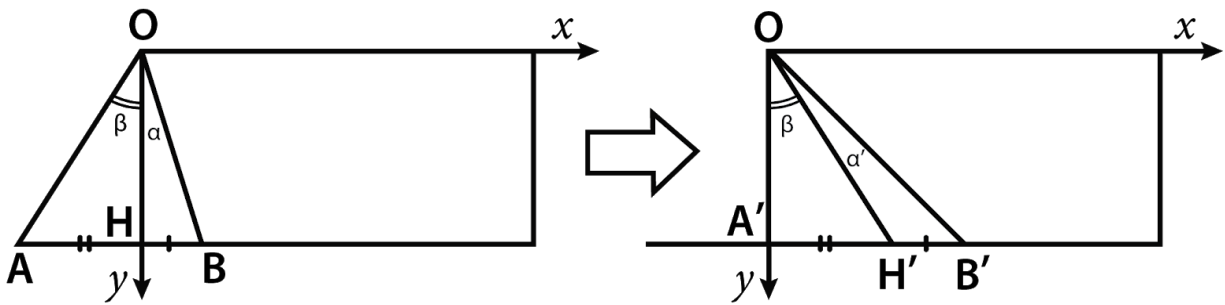

Figure 4 - example of shear by angle : point A changes into A', H into H', B into B'.

Note that no cyclic effect will occur even after the horizontal shear is applied because we have extended the original image: the «black zone» will simply move and change its from (see figure 5).

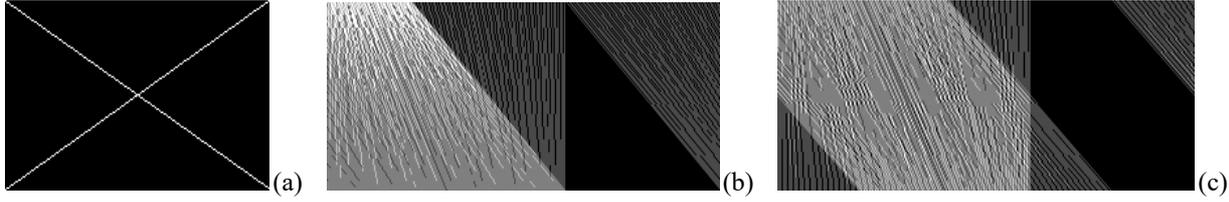

Figure 5 – (a) image example with the white cross; (b) Hough transform for quadrant $[315°; 0°]$; (c) Hough transform after horizontal shear by $arctg(\frac{1}{2})$.

It is now clear that we can use horizontal shear to transform any angle range $[-\gamma 1; \gamma 2]$ into $[-arctg(tg(\gamma 1) + tg(\gamma 2)); 0]$ and then apply scale to transform the result into $[315°; 0°]$. If the original range was formed by $\gamma 1 = \gamma 2 = arctg(\frac{1}{2})$ the result of these two transformations will be $\gamma$-neighbourhood of zero angle.

Now let's consider how to perform Hough transform simultaneously with the two operations described above (since additional operations will increase the duration and the computational complexity of our task). For horizontal shear by shear angle $\gamma$ we introduce parameter *spl* (*shift per line*) $= tg(\gamma)$. To perform the shear each row of the original image must be shifted by *spl*. In [2] rows are calculated in the following way: for each iteration Hough transform for "mostly vertical" lines is calculated for all horizontal blocks calculated on the result of the previous iteration (see figure 6).

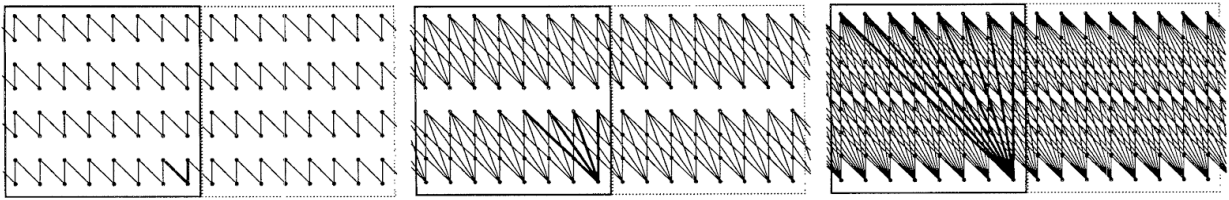

Figure 6 - iterations of FHT algorithm

Let's change the algorithm so that on the first iteration rows are calculated as if they are already (cyclically) shifted by *spl* (all other iterations remain the same). Optimization of vector operations on modern processors allows to simulate horizontal shear of the image during the transform calculations without additional memory use or performing additional operations. In case of scale the approach is the same - on the first iteration we scale rows of the original image. Additional resources are required only in case of dilation - we will need a fixed amount of additional memory.

## 5. FHTSHIFT

Note that visual representation of Hough image is not always easily understandable due to the lines "flowing through" the boundaries of Hough-image. We can perform an additional post-processing of Hough-image (we call it *fhtshift*) to get rid of this effect - i.e. so that moving from left to right on any row we will be passing through the points that are images of lines that are all at the same angle and that are located on the original image from left to right. In every row the number of "meaningful" pixels will be different and equal to $w + h * tg(\alpha)$ (see figure 7b), changing from $w$ for vertical line (at angle $0°$) to $w + h$ or line at $45°$ angle.

This can be achieved by shifting cyclically each row based on the angle range of the respective quadrant. For example, for quadrant $[315°; 0°]$ its «black zone» is a triangle and its edges are images of points $(0, h)$, $(w, 0)$ on the original image (remember that here we work with the flipped image). Their coordinates in Hough-image will be $[(w, 0) \rightarrow (w, h)]$ and $[(w, 0) \rightarrow (w + h, h)]$.

To achieve the desired result we need to move $[(w + h, 0) \rightarrow (w + \frac{h}{2}, h)]$ and $[(0, 0) \rightarrow (\frac{h}{2}, h)]$. This means that row zero will shift by $h$ and the last row (where *shift* $= h$) will shift by $\frac{h}{2}$. Rows are shifted one-by-one by $h - \frac{i}{2}$, where $i$ is the row number (see figure 7).

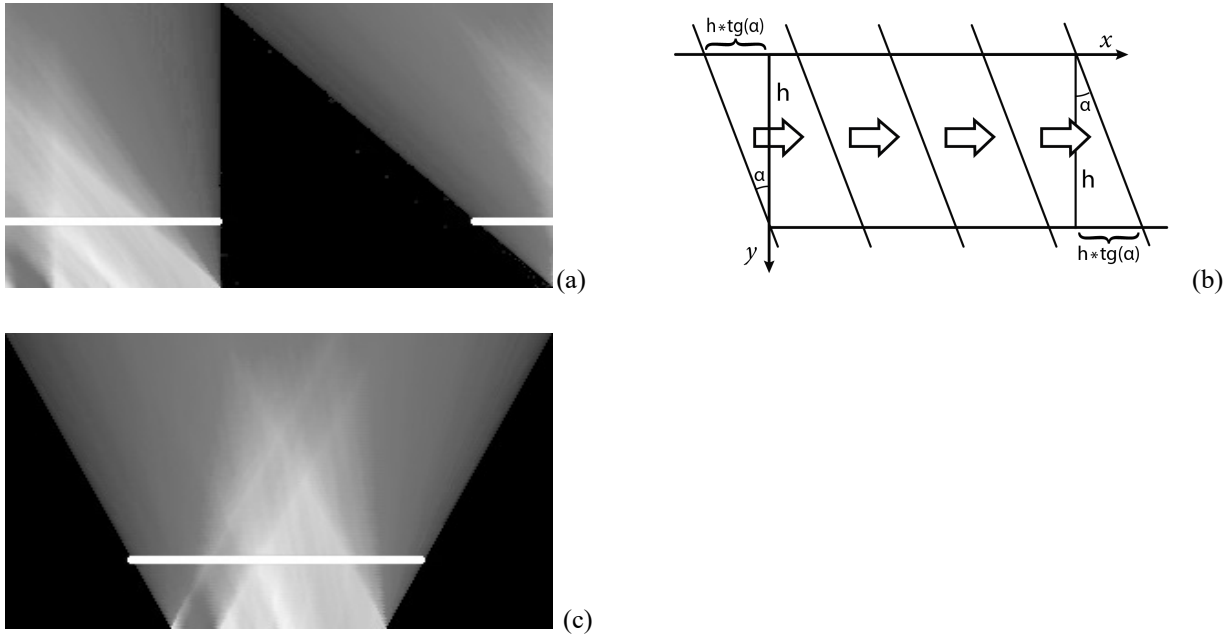

Figure 7 - (a) Hough transform result for quadrant $[315°; 0°]$; (b) illustrative example of lines with images on one row, and (c) quadrant $[315°; 0°]$ after post-processing.

The following image will be the result of full Hough transform (see figure 8):

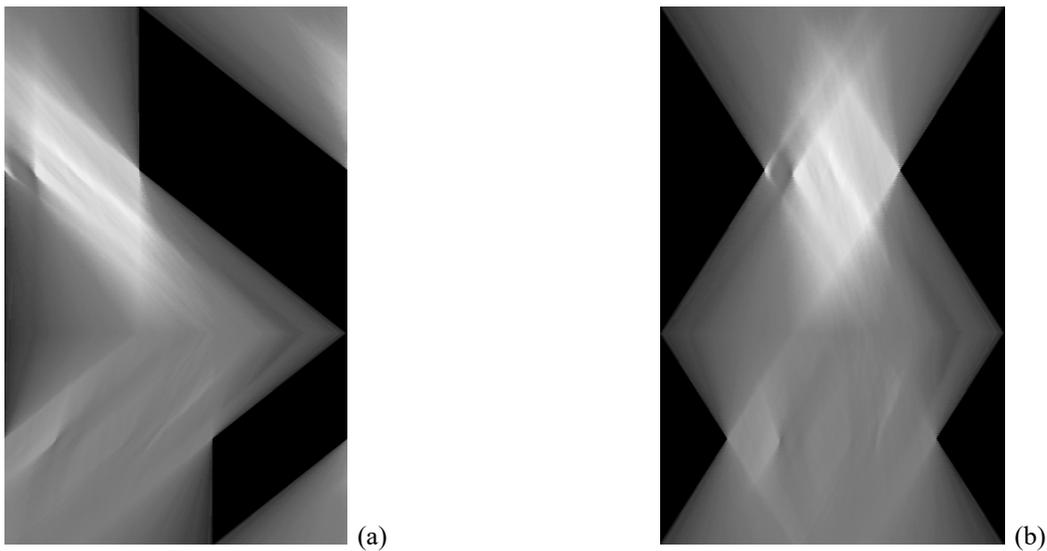

Figure 8 - (a) result of full Hough transform before and (b) after post-processing.

To better understand *fhtshift* we can use as an example fftshift function that is used for Fourier transforms. Before applying *fhtshift* the "black zone" location on Hough-image depends on the properties of the original image. After *fhtshift* the entire "black zone" will be on the edges of Hough-image and all its "meaningful" pixels will be concentrated around its center in a single zone without any gaps.

## 6. CONCLUSION

As we have shown in section 3.3 any point in concatenated quadrants of only "mostly vertical" (or only "mostly horizontal") lines on the original image maps to a line in Hough-image as a result of Hough transform. Since full Hough transform is the concatenation of these two (concatenated only "mostly vertical" quadrants and concatenated only "mostly horizontal" quadrants), the image of a point on full Hough image will be two linear patterns with the common point on the common row after concatenation. Therefore, in described realization of Hough transform every point in original image corresponds in Hough-image to a figure consisting of two segments ("angle"). Proposed realization of Hough transform is as close as possible to "ideal" point-to-line mapping that transforms bounded area (original image) into bounded area (Hough image).

In this work we present algorithm modification for calculating fast Hough transform to obtaining sums along lines on the original image for a defined range of inclination angles. This modification allows to decrease computational complexity compared to computing full Hough-image.

For clear visualisation of Hough-image new operation "*fhtshift*" is introduced, which regroups "meaningful" part of Hough-image to simply connected region.

## ACKNOWLEDGMENTS

This work is partially supported by Russian Foundation for Basic Research (projects 17-29-03240, 17-29-03297).